# Fabrication Of Economical Signal Amplified Vibrating Sample Magnetometer With Spiral Designed Detector

Aditya Dash[1,a)], Rabindra Sarkar[1,b)], Abel Mathew[1,c)], Prakash Nath Vishwakarma[1,d)*]

[1]*Department of Physics and Astronomy, National Institute of Technology Rourkela, Odisha, India.*

a) adityadashmay@gmail.com
b) 416ph5052@nitrkl.ac.in
c) 416ph5047@nitrkl.ac.in
d) ***Corresponding Author**: prakashn@nitrkl.ac.in

**Abstract.** The design and fabrication of a cost-effective vibrating sample magnetometer are explained. Improvement in the detected signal can be obtained using an operational amplifier circuit. The fabricated spiral detection coil design enhances the induced voltage obtained as shielding flux is reduced. A homemade setup is obtained by taking a woofer as an actuator and plexiglass as the vibrating medium. Lock-in amplifier analyzed the booster enhanced induced voltage signal by the principle of phase detection. Study of amplified and non amplified signals from the Nickel (99% purity) sample was done. Conversion of induced voltage to magnetization was done by calibration incorporating the coercivity and retentivity measurement. Magnetic oxide can be analyzed using this cost-effective designed Vibrating sample magnetometer. The data obtained from the VSM can conclude the successful operation of the designed magnetometer.

**Keywords**. Vibrating sample magnetometer, Operational amplifier, Lock-in Amplifier, Coercitivity, Retentivity, Saturation Magnetisation, Actuator.

## 1. Introduction

In today's era of the electronic uprising, magnetic measurements are the most revolutionary in the development of nanochips and integrated circuits. Improvement in the instrumentation part of the technologies is a necessary procedure to examine any magnetic or non-magnetic samples. Vibrating Sample Magnetometer acts as a boon for the magnetic measurement and determining the M-H curve[1]. However, the progression should primarily include the low cost, enhanced sensitivity, speed of operation and accuracy. Use of various methods and instruments, to measure the magnetization of the sample is available. Broadly, the magnetization measurement techniques can be classified into three categories. Firstly, force measurement in a non-uniform magnetic field technique and then, the magnetic induction measurement technique. And finally, the measurement technique of indirect phenomenon involving magnetization. Vibrating Sample Magnetometer (VSM) belongs to the magnetic induction measurement technique[2]. This induction based measurement technique consists of a field coil, that is used to magnetize the sample and magnetization is measured as the induced voltage observed by the pick-up coils. The induced voltage is produced due to change in the position of the magnetized sample or the coil, which leads to a change in magnetic flux with respect to the movement of the sample[3]. And the relative change in flux produces induced voltage following the principle of Faraday's law of induction in the coils, either the sample can move or the coils[4]. The techniques, which involve the progressive movement of coils, are generally cumbersome and very difficult to control. The main drawback is that it requires an extremely uniform magnetic field, and the effects caused due to the non-uniformity are very difficult to correct [5]. In addition, in a uniform magnetic field, huge corrections are required for the samples with very small magnetic moments. So, the magnetometers with static coils and moving sample are the competent candidate among the other types of vibrating sample magnetometer. Here we are discussing a cost-effective and successful way to construct a Vibrating Sample Magnetometer with a higher order of accuracy[6]. A modern and effective design of the detection coil has been made, which is the prominent part in terms of sensitivity of the magnetometer. Our main purpose is to amplify the induced signal using a booster circuit[10]. Our setup consists of a speaker, four detection coils, a sample holder, a laboratory electromagnet and lock-in-amplifier. An electromagnet is used to magnetized the sample, a speaker is used to oscillate the sample, and four detection loops are arranged in a way to increase the sensitivity. Relative motion between the sample and the coil produces an induced EMF and the pick-up coils detect the EMF generated by the motion. Then the output of the coils is fed into the lock-in-amplifier for measurement. The speaker is given input from the lock-in-amplifier through which the amplitude and the frequency of the vibration can be controlled. Further, many small modifications are made to amplify the output signal and accurately measure the magnetization of the sample. In our study, we have analyzed the results of two samples (Nickel and mild steel), which are used as samples, before and after adding the signal amplifier unit.

## 2. Experimental

**Mechanical design of the vibrating sample magnetometer**

A woofer or a bass speaker having impedance 4 Ω was chosen for the periodic, harmonic and approximately one-dimensional oscillation (vibration) as the replacement of the vibrational driver of the VSM It works on 10W, 0-5 V power supply. The woofer can work as a linear actuator for a particular axis vibration. A phenolic fabric cap (hylam) was attached to the polysynthetic diaphragm of the woofer using adhesives for the stability of the system. Plexiglass rod acts as a vibrating medium for the sample. It is then attached to the hylam cap with proper orientation and verifying the fixed amplitude of the oscillation. A small cut out is made in the lower portion of the rod for holding the sample. This will incorporate the sample, which is then glued to the holder in an orientation either perpendicular or parallel to the magnetic field lines. The whole vibrational set up was assembled in a wooden frame with four rods which is then screwed to the base for stability. The height of the woofer can be adjusted so that the adjustments in the sample position along the vibrational to axis can be made. The spacings between the four levelling screws are 52.69 mm, 52.32 mm, 55.95 mm, 51.76 mm, respectively. The woofer must have a clearance of 20 mm from the base in order to reduce the risk of noise in the system. The four-rod system must have sponge material at the junction in order to reduce the vibrational noise which can be seen in Fig 1.

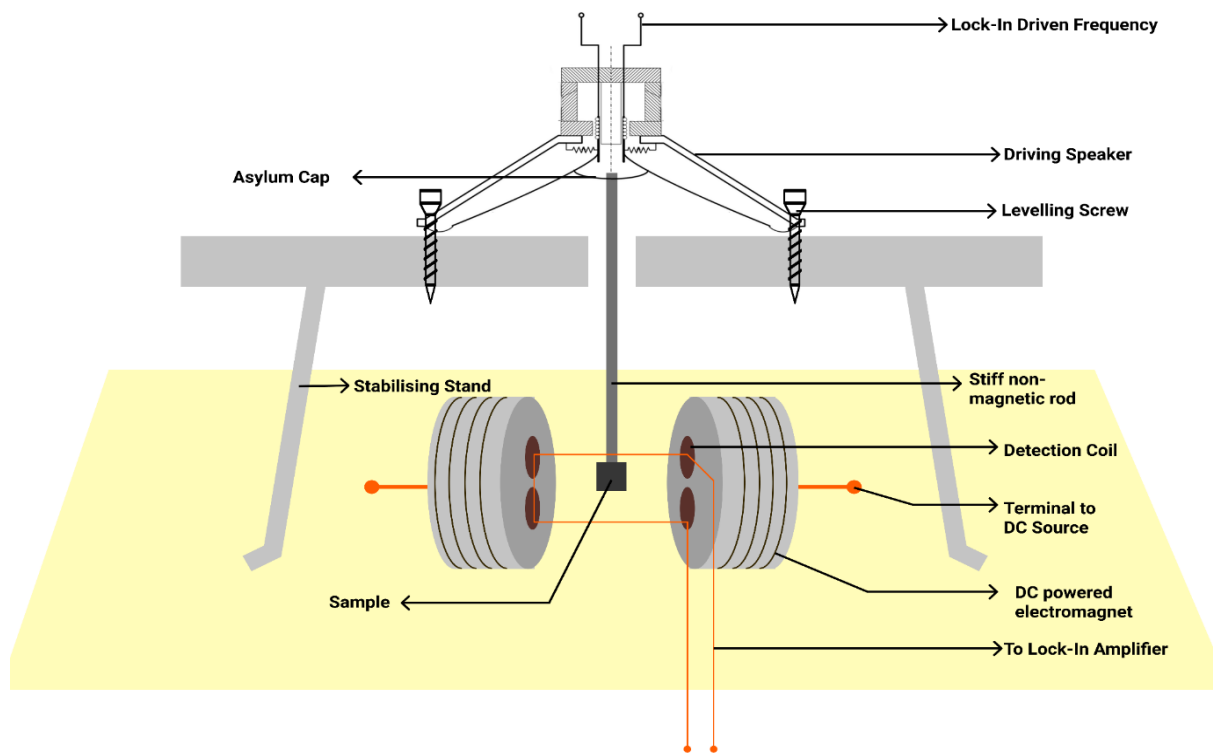

**Figure 1**: Schematics of homemade vibrating sample magnetometer (VSM) set up

**Electromagnet set up for VSM**

A laboratory electromagnet for producing uniform parallel magnetic field was used having specification SES Instruments model #EMU-50 as shown in Fig 2(c). A voltage-controlled DC current source ( 0 – 4 Ampere) used for the generation of the magnetic field. The magnetic field was calibrated using a gaussmeter for different DC current inputs. The detection coils were attached to the poles by adhesives, and a minimum separation of 15.47 mm is maintained between the poles for the maximum magnetic field between the pole pieces.

**Detection coils**

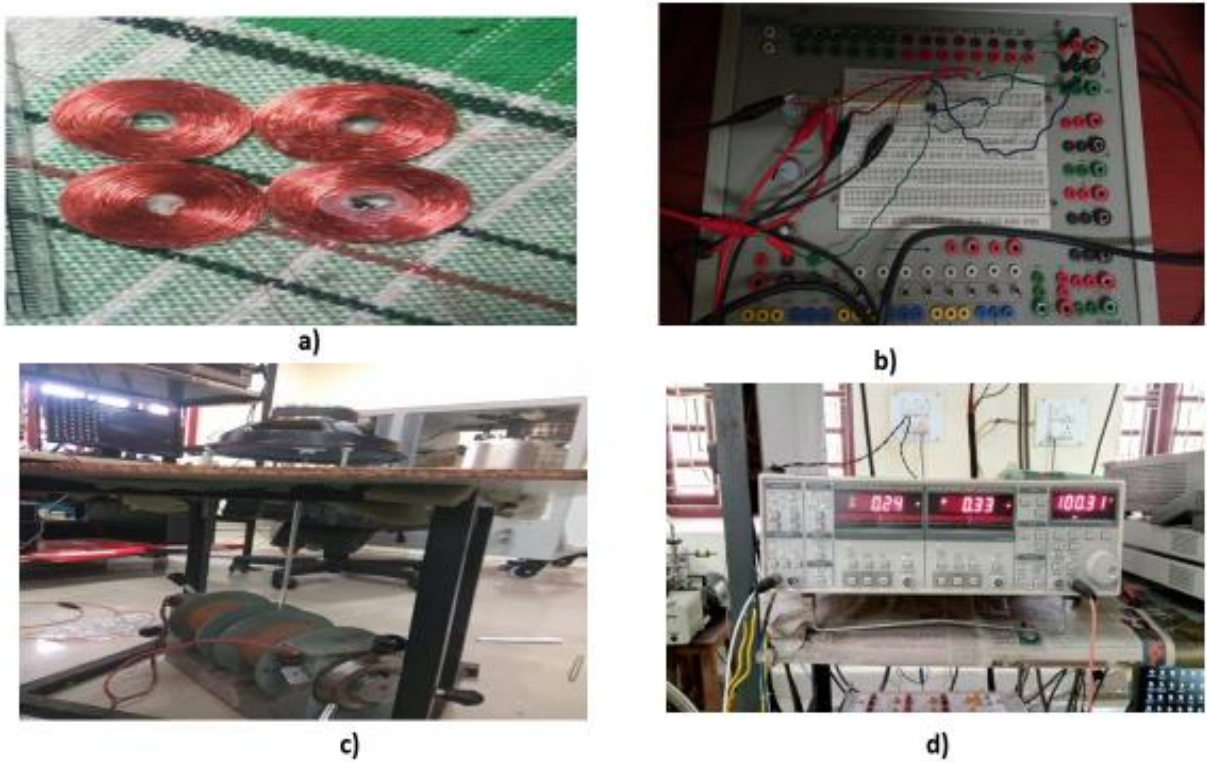

**Figure 2**: a)Detection coils used in vibrating sample magnetometer (VSM) set up b) Snap of the non-inverting Amplifier for booster circuit c) Homemade Vibrating sample magnetometer set up d) Stanford model SR830 Lock-in amplifier

The induced voltage developed and the sensitivity depends primarily on the detection coil design as in Fig 2(a). A thin copper wire was used and spiralled to produce thin sheet spiral structure. The diameter of the spiral coil was 1.98 cm incorporating nearly 200 numbers of turn. Shielding effect due to copper bulk diamagnetic behaviour is reduced in spiral arrangement as piling up is less. Minimum shielding flux loss leads to enhanced sensitivity.

**Sample and amplification design**

Measurement of the hysteresis loop was done for nickel sample and mild steel (1.2% carbon content) sample. The weight and dimension of Nickel (99% pure) was 175.69 mg and 0.165 mm*5.34 mm*11.42 mm, respectively. Mild steel had 202.5 mg as weight and 0.192mm*6.84 mm*10.92 mm as dimension.

The non-inverting amplifier circuit was designed for small ac signal to amplify the voltage induced, as shown in Fig 2(b). IC 741 opamp was used with +Vcc= 9V and –Vcc= -9V and $R_2$=10 k Ω, $R_1$= 1k Ω. One end of the coil was fed to non-inverting end, and another end is grounded. Hysteresis loop was obtained for the Nickel.

**Lock-in Amplifier**

Lockin amplifier (SR830 DSP Lock-in amplifier Stanford Model ) supply the power to the woofer (vibrating system) with a given input frequency of 550.7 Hz and 4.5 V supply as shown in Fig 2(d). The time constant was fixed at 30 seconds, and sensitivity was locked at 200 mV [7].

## 3. Results and discussion

Hysteresis loop was obtained from the induced voltage developed in the spiral coils varying with the magnetic field intensity. By Faraday's law of electromagnetic induction, flux φ linked to the spiral coils is $\mu\mu_0 H$ and the voltage induced $V(t) = -N \, \partial\varphi(t)/\partial t$ where H is the magnetic field and N is the number of turns in spiral coil [11][12]. The calibration of the magnetic field with respect to the input current from the constant DC current source was done from -4 A to 4 A input range shown in Fig 3(a). The signal induced in the spiral coil depends upon the frequency of vibration fed by the lock-in amplifier. By analyzing the induced voltage at different frequencies for constant driving voltage of 4.5 V, we got a peak voltage for a frequency of 550.7 Hz. So the overall process was done for the optimized working frequency. Sinusoidal voltage is supplied to the vibrational driver, using the lock-in amplifier. Detection of the induced voltage (at the locked frequency) was done by the lock-in amplifier. The lock-in amplifier works on the principle of phase-sensitive detection to filter out the single-frequency output out of all residue noises. The time constant signifies the averaging out of the fluctuating induced voltage over a time period which was chosen as 30 seconds.

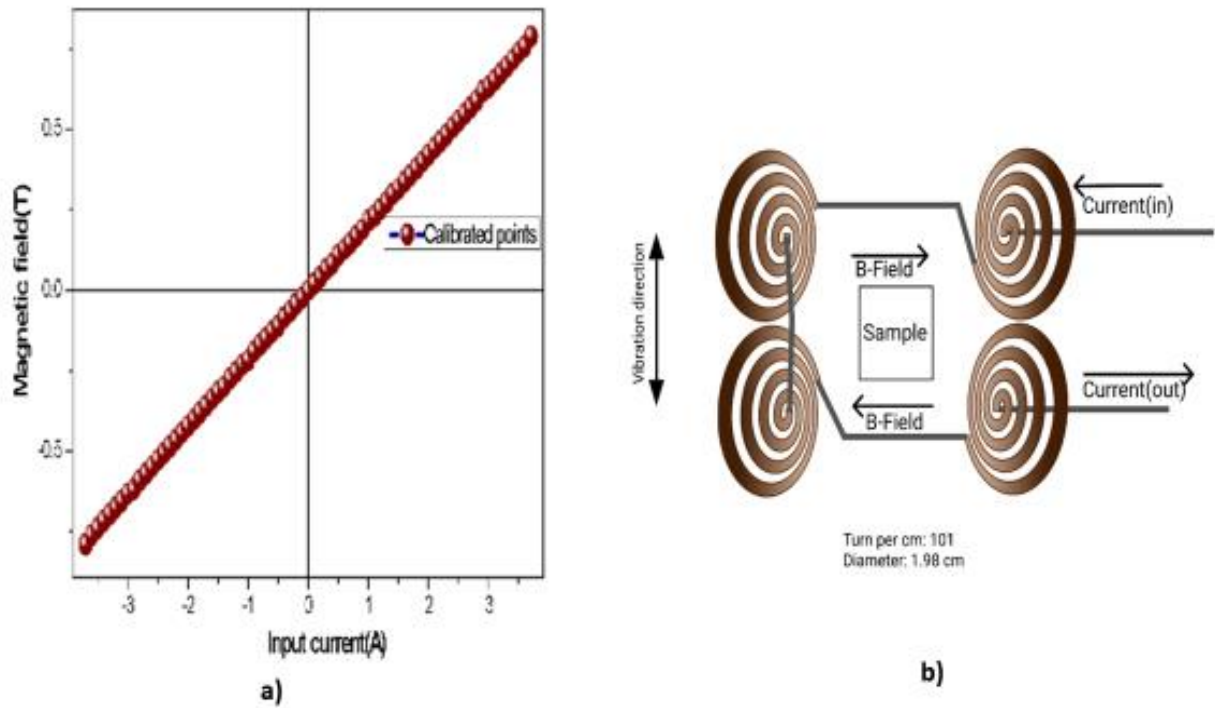

**Figure 3**: **a)** Variation of the magnetic field with DC current supplied for all values of current -4 A to 4 A  **b)** Spiral detection coils and its specified orientation to obtain maximum induced voltage

The copper enamelled thin spiral detecting coils was the main component which affects the sensitivity of the Vibrating Sample Magnetometer. The current flow through the two coils attached in each pole in such a way that the magnetic field developed as a result of this current, is in the opposite direction according to the right-hand thumb rule. EMFs of opposite polarization in S1+S2 and S3+ S4 are induced as in Fig 3(b). Since the coils are connected in series opposition, the two signals are added up [8]. The clockwise or anticlockwise spiralling of the coils is to be taken into consideration while connecting. These coils were in series opposite resistance so that the residual magnetic field from outside to be cancelled out and the voltage induced to be summed up as shown in fig 6. The sample should be adjusted so that the vibration amplitude was linked to both the coils in series. Higher the number of turns more will be the induced voltage. The advantages of using this coil configuration are that it has reduced effective shielding as compared to other configuration and overall resistance of the coils is low, so enhanced sensitivity is expected.

The sample placing has some serious problems which need to be looked upon. The sample should have a clearance distance between 2 poles. Otherwise, it will get magnetized and attached to the poles, thereby dampening the vibrations. The tip end of the sample holder should be light. Otherwise, it will affect the unidirectional vibration. Degaussing of the sample is the major requirement if the sample is already subjected to the magnetic field. Chronologically increasing and decreasing stepwise magnetic field will work or we have to heat it to a very high temperature that is the Curie temperature which makes it into the original configuration with no stored magnetic field. For the nickel sample, we have directly obtained the data points in terms of induced voltage for a given applied DC current (magnetic field intensity) to the electromagnet as shown in Fig 4(a).

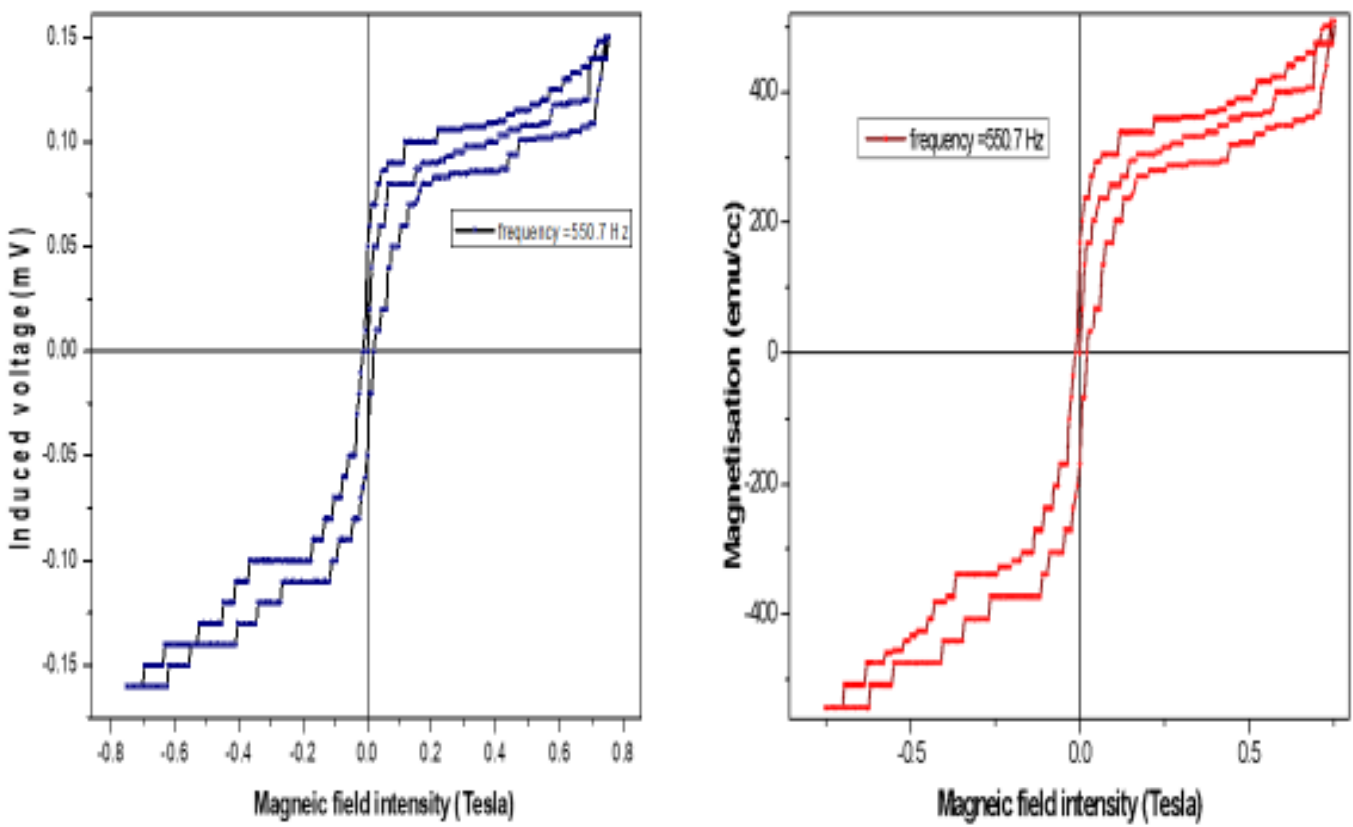

**Figure 4**: (a)Variation of induced voltage with respect to the magnetic field intensity for Nickel (99%) (b)Variation of magnetization with respect to the magnetic field intensity for Nickel (99%)

Before going into the measurement part of magnetization, the induced voltage to be calibrated with respect to the magnetization. The density of Nickel is around 8.902 g/cm$^3$, and the saturation magnetization was calculated as for the Nickel sample[9]

$$M_s = (0.60\ \mu B/atom) \times (9.27 \times 10^{-24}\ Am^2/\mu B) \times (0.9135 \times 10^{29}\ atom/m^3)$$
$$= 5.08 \times 10^5\ A/m$$
$$= 508\ emu/cm^3$$

The constant was

$$CO = \left(\frac{Vsaturation}{Msaturation}\right) \times \frac{1}{Volume}$$

$$= \left(\frac{0.156 mV}{508\ emu/cm^3}\right) \times \frac{1}{10.06 \times 10^{-3} cm^3}$$

$$= 0.03052\ mV/emu$$

The constant CO was used to calibrate to get the magnetization variation with magnetic field intensity as mentioned in the Fig 4(b). Hysteresis loop was obtained for the Nickel (99%) in terms of induced voltage and magnetizing field intensity to the electromagnet. The saturation induced voltage nears 0.156 mV and retentivity induced voltage nears 0.056 mV.

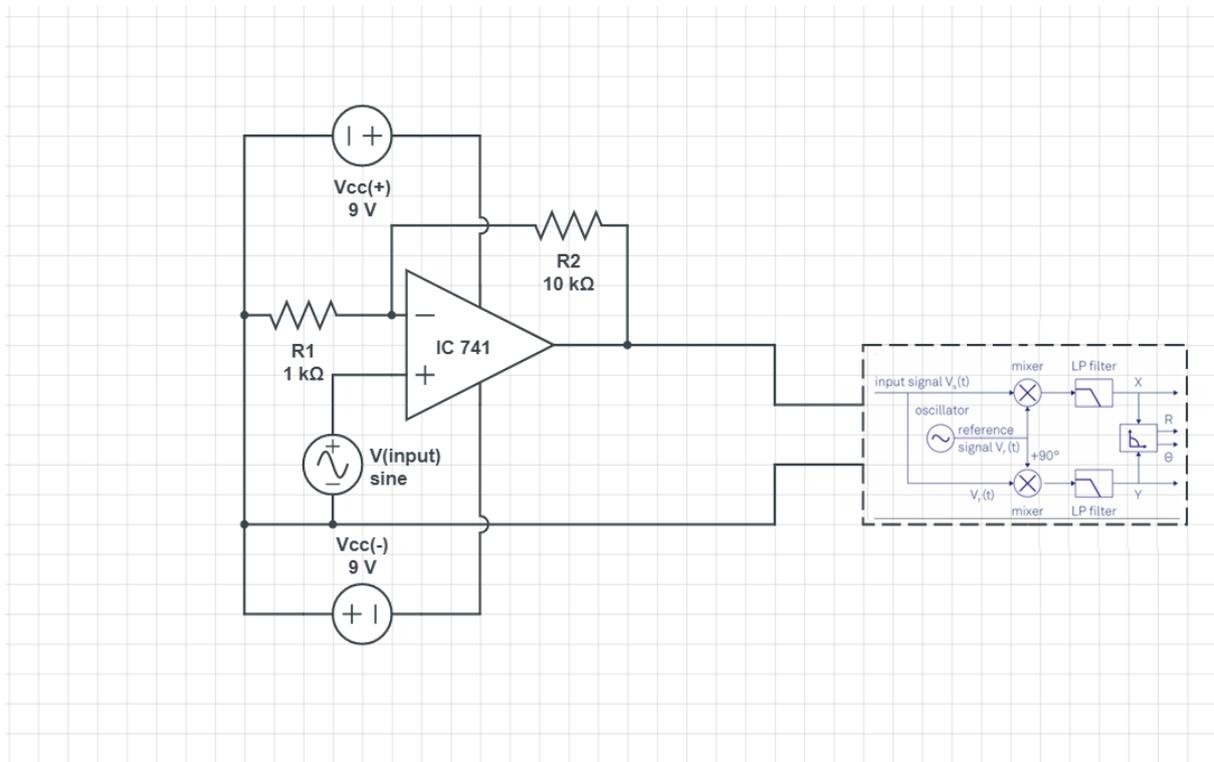

**Figure 5**: Non inverting Amplifier set up using Op-amp and lock-in amplifier phase circuit (the output of VSM is fed to the amplifier)

The coercivity of the Nickel (99%) can be interpreted from the magnetization curve is -0.013 T (10.4 kA/m). The retentivity obtained from the first quadrant of the magnetization curve is 169.26 emu/cc. The data obtained matches with the previous study of Nickel magnetization analysis. An operational amplifier circuit is implemented to amplify the output signal before feeding into the lock-in amplifier. Although the lock-in can detect very small signals, the amplification of induced voltage before feeding to lock-in amplifier is done to increase the sensitivity of the signal. This non-inverting amplifier act as booster circuit, whose output is in phase with the input. The circuit configuration is shown in Fig 5. IC 741 is used, which is a high performance and a high open-loop gain operational amplifier. The output from the VSM is fed to the non-inverting input of the opamp. The inverting end is connected to $R_1$(input resistance), which is grounded. $R_2$ is the feedback resistor which in together with the $R_1$, determines the gain. The +Vcc and –Vcc are the power supply voltages on which the opamp works. The output of the amplifier is fed to the lock-in amplifier.

The open-loop gain can be calculated as

$$Av = 1 + \frac{R_2}{R_1}$$

$$= 1 + \frac{10}{1}$$

$$= 11$$

For a particular current value 3.48 A, we will calculate the observed gain in the induced voltage.

$V_{in}$ = 0.146 mV

$V_{out}$ = 1.226 mV

$$Ao = \frac{1.226}{0.146}$$

$$= 8.397$$

We have amplified the induced voltage using the booster circuit for the Ni sample. The variation of induced voltage with the magnetic field intensity can be interpreted from Fig 6(a).

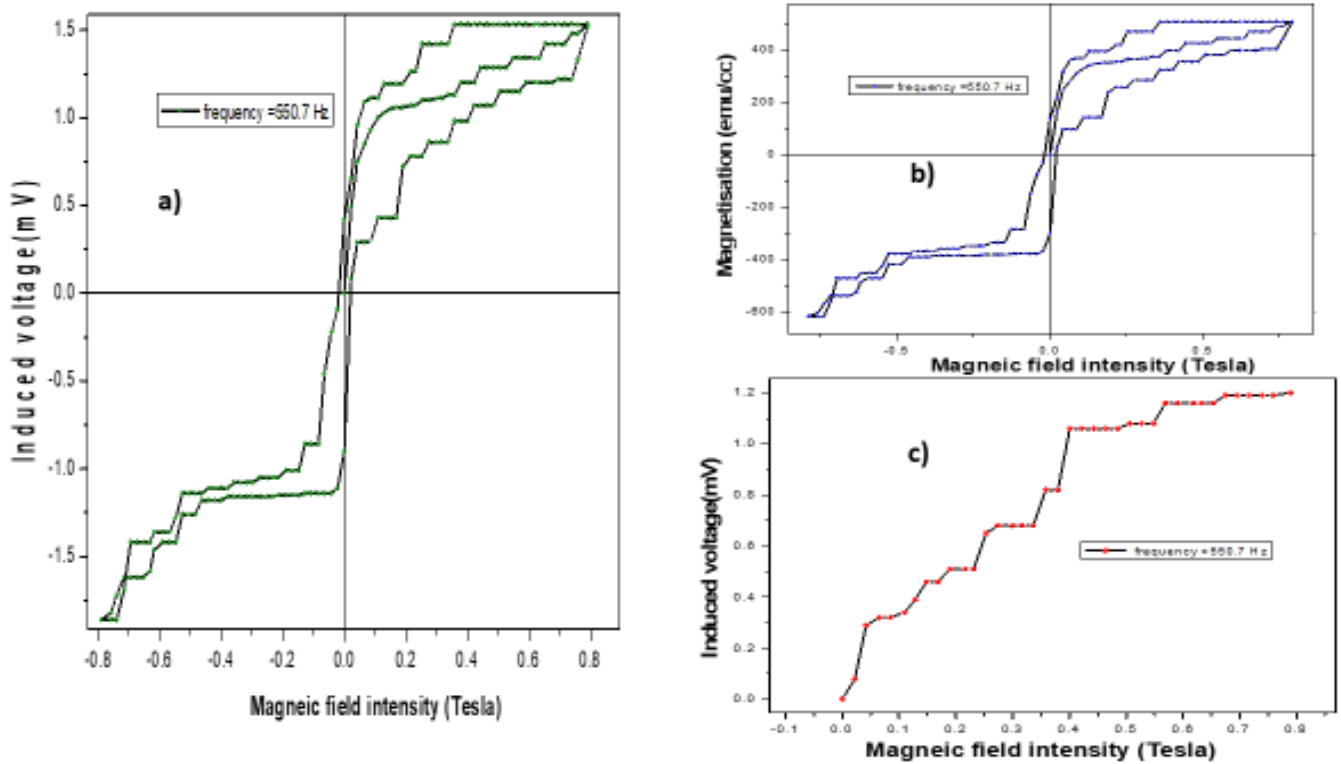

**Figure 6**: (a) Variation of induced voltage with respect to the magnetic field intensity for Nickel (99%) using an amplifier circuit. (b) Variation of magnetization with magnetic field intensity for Nickel using amplifier circuit. (c) Variation of induced voltage with respect to the magnetic field intensity for mild steel.

Hysteresis loop for the amplified signal of the same nickel sample was studied. The saturation induced voltage nears 1.33 mV, and retentivity induced voltage nears 0.42 mV. For amplified voltage, the calibration constant can be calculated as

$$CO = \left(\frac{Vsaturation}{Msaturation}\right) \times \frac{1}{Volume}$$

$$= \left(\frac{1.336 mV}{508\ emu/cm^3}\right) \times \frac{1}{10.06 \times 10^{-3} cm^3}$$

$$= 0.2614\ mv/emu$$

The induced signal will be divided by this factor and volume to obtain the magnetization. The obtained magnetization curve can be analyzed further. The hysteresis loop obtained was analyzed to obtain the different magnetic parameters as in Fig 6(b). The coercivity of the sample is around -0.018T (14.4 kA/m). The obtained retentivity from the plot is 139.27 emu/cc. Magnetic moment study for the mild steel sample was tried. But the saturation magnetization for the mild steel was out of the DC current source range. Even the electromagnet heats up if we increase the DC current further. The heating affects the magnetic field generation also caused the detection coil to get detached from the pole. A high magnetic field intensity was required to get to the saturation region. The plot up to which the DC current or the magnetic field intensity can be maximized is in Fig 6 (c).

**Further scope**

Temperature variation set up can be incorporated by using nichrome heater wire set up for high temperature and liquid nitrogen set up for low-temperature variation. We need a high magnetic field electromagnet to get the hysteresis loop for high saturation magnetization material and low magnetic material. And the lock-in amplifiers are very costly to use. So in order to replace it, we need a phase-sensitive detector and a bandpass filter to filter out the noise. To get an inert condition around the sample, a vacuum set up is required. Piezoelectric actuators can be used to increase the amplitude of vibrations which itself increases the sensitivity of VSM In order to get automated data readings, DAQ can be incorporated by using Labview programs.

## 4. Conclusion

A cost-effective vibrating sample magnetometer is designed and fabricated using an actuator (woofer) and a lock-in amplifier as a primary component. The designed detection coil has a lot of superiority as compared to other primitive pick-up coils. These coils have lesser shielding flux loss and more sensing capabilities. The orientation of all the pick-up coil is such that the generated EMF in each coil will add up to give maximum output as a signal. We have designed an operational amplifier circuit which can act as a booster circuit for the induced signal amplification from coils. Analysis of non-amplified and amplified induced voltage was done for the Nickel (99%) sample. Lock-in amplifier was not able to analyze very low signal properly. So we have used a booster circuit to amplify the signal before feeding into the lock-in amplifier. Conversion of the induced voltage to the magnetization was done using a simple calibration technique. The coercivity and retentivity obtained for the nickel (99%) sample both for amplified and non - amplified loop is in accordance with the previous experimental analysis and literature reported for the Nickel. For the mild steel sample, sufficient magnetizing field intensity was beyond the operating range of the DC current source, which provides the current to the electromagnet. The saturation magnetization can be obtained at the very high magnetic field for the mild steel because of its high content of Iron on it. We are hoping that some ferromagnetic oxides and other samples can also be analyzed with this cost-effective vibrating sample magnetometer. From the data obtained, we can conclude the successful operation of the fabricated VSM.

## 5. Acknowledgements